\documentclass[%
 aip,
 apl,%
 amsmath,amssymb,
 reprint,%
]{revtex4-1}

\usepackage{graphicx}
\usepackage{dcolumn}
\usepackage{bm}

\newcommand{\be}{\begin{equation}}
\newcommand{\ee}{\end{equation}}

\usepackage{color}

\begin{document}

\title{Material dependence of Casimir forces: gradient expansion beyond proximity}

\author{Giuseppe Bimonte}
\affiliation{Dipartimento di Scienze
Fisiche, Universit{\`a} di Napoli Federico II, Complesso Universitario
MSA, Via Cintia, I-80126 Napoli, Italy}
\affiliation{INFN Sezione di Napoli, I-80126 Napoli, Italy}

\author{Thorsten Emig}
\affiliation{LPTMS, CNRS UMR 8626, B\^at.~100,
Universit\'e Paris-Sud, 91405 Orsay cedex, France}

\author{Mehran Kardar}
\affiliation{Massachusetts Institute of
Technology, Department of Physics, Cambridge, Massachusetts 02139,~USA}

\date{\today}

\begin{abstract}
  A widely used method for estimating Casimir interactions
  [H.~B.~G. Casimir, Proc. K. Ned. Akad. Wet. {\bf 51}, 793 (1948)]
  between gently curved material surfaces at short distances is the
  proximity force approximation (PFA). While this approximation is
  asymptotically exact at vanishing separations, quantifying
  corrections to PFA has been notoriously difficult. Here we use a
  derivative expansion to compute the leading curvature correction to
  PFA for metals (gold) at room temperature. We derive an explicit
  expression for the amplitude $\hat\theta_1$ of the PFA correction to
  the force gradient for axially symmetric surfaces.  In the
  non-retarded limit, the corrections to the Casimir free energy are
  found to scale logarithmically with distance.  For gold,
  $\hat\theta_1$ has an unusually large temperature dependence.
\end{abstract}


\maketitle

The continual drive towards miniaturization of devices inevitably leads to scales were quantum effects
are significant. In the realm of micromechanical systems (MEMS)--
nano fabricated devices actuated by electrical bias-- Casimir\cite{Casimir48} and van der Waals
forces\cite{Parsegian} are paramount.
These forces originate in quantum fluctuations of electromagnetic fields, and may cause devices to
fail due to  stiction.\cite{MEMS}
Due to their non-trivial dependence on shape and material properties, Casimir forces are difficult
to compute.\cite{Johnson,worldline}
A common  method for estimating forces in a general setup  is the proximity force approximation
(PFA):
First the force is calculated between two parallel plates at separation $d$ with the Lifshitz
formula (as a function of dielectric response of the adjoining slabs;\cite{Lifshitz})
then the geometry is taken into account by averaging over (appropriately defined) separations $d$
between adjoining surfaces.\cite{Derjaguin}
While this approximation is qualitatively wrong in special circumstances,\cite{Alejandro}
it remains a useful tool in high precision experiments between surfaces of large radii of curvature $R$.
In these cases PFA is asymptotically exact at small separations (for $d\ll R$).
However, improvements in sensitivity of measurements warrant quantifying corrections to PFA
which we undertake in this paper.

\begin{figure}
\includegraphics[width=.9\columnwidth]{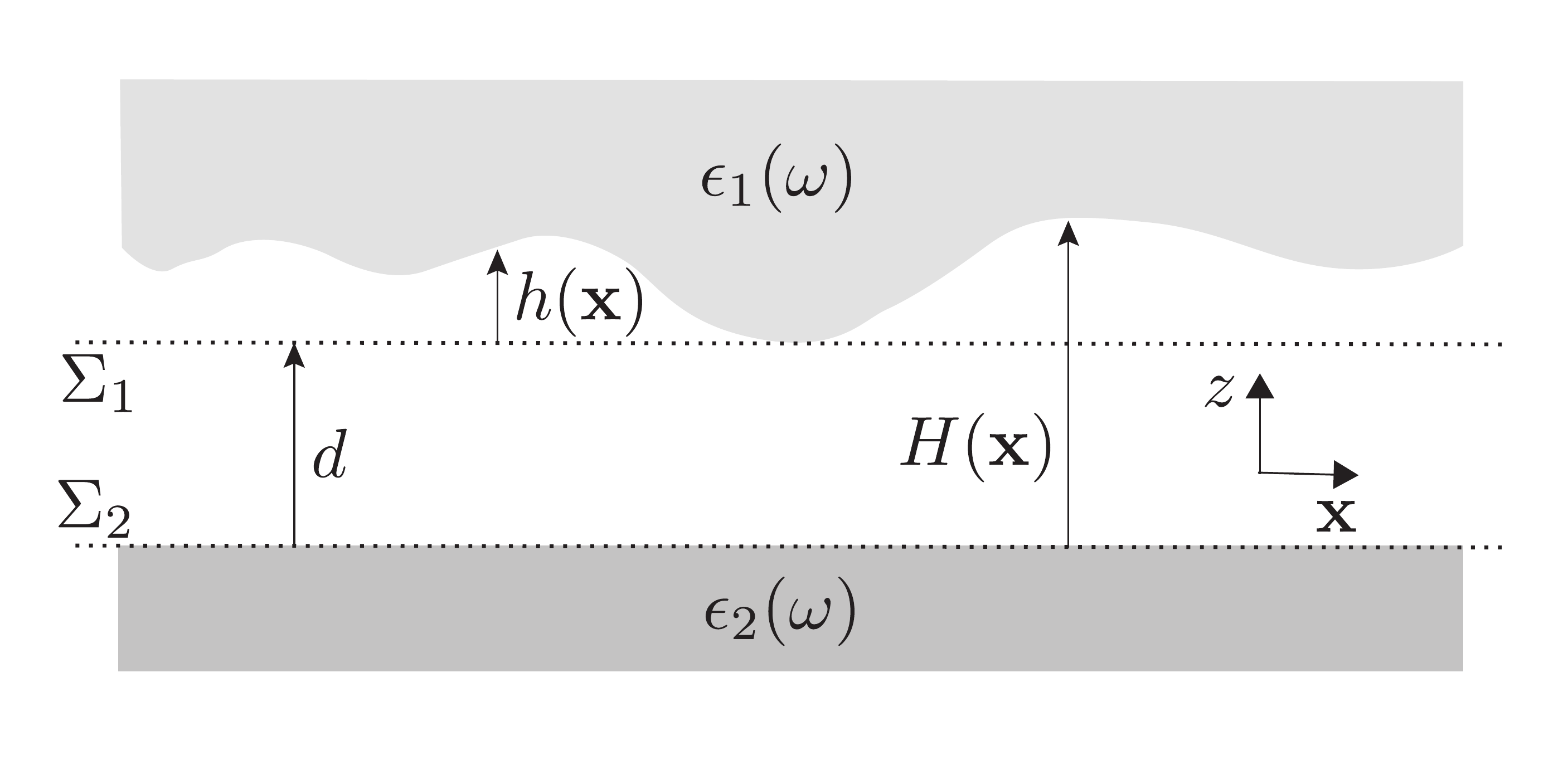}
\caption{\label{fig1} Parametrization of a the profile of a gently curved dielectric surface near
a flat dielectric plate.}
\end{figure}

We consider a geometry consisting of two infinitely thick plates composed of homogeneous and isotropic dielectric materials, with  permittivities $\epsilon_1(\omega)$ and $\epsilon_2(\omega)$. For simplicity we assume that one of the plates is a plane-parallel slab, while the other    is gently curved, and characterized by a {\it smooth} height profile $z=H({\bf x})$, where $z$ is the local distance from the planar surface $\Sigma_2$, and  ${\bf x} =(x_1,x_2)$ is the vector spanning  $\Sigma_2$ (see Fig.~\ref{fig1}).\cite{fn1} Following Refs.~\onlinecite{fosco,Bimonte2011} we postulate that the Casimir free energy ${\cal F}[H]$  admits a {\it local} expansion  of the form
\begin{equation}
\label{eq.1}
{\cal F}[H]={\cal F}_{\rm PFA}[H]+\int_{\Sigma_2} d{\bf x}\,   \alpha(H)
{\bf \nabla} H \cdot {\bf \nabla} H  +  { \rho}^{(2)}[H]\;,
\end{equation}
where ${\cal F}_{\rm PFA}[H]$ represents the PFA free energy
\begin{equation}
{\cal F}_{\rm PFA}[H]=\int_{\Sigma_2} d{\bf x}\, {\cal F}_{pp}(H)\;,
\label{eq.100}
\end{equation}
with ${\cal F}_{pp}(z)$ the free energy per unit area for two plane-parallel dielectric plates of permittivities $\epsilon_1(\omega)$ and $\epsilon_2(\omega)$ at distance $z$, as given by the Lifshitz formula,\cite{Lifshitz} and $\alpha(H)$ is a function to be determined.  The quantity $ {\rho}^{(2)}[H]$ in Eq.~(\ref{eq.1}) represents corrections
that become negligible if the curvature of the surface is small compared to the minimal distance $d$ between the surfaces.\cite{fn2}

The function $\alpha(H)$ in Eq.~(\ref{eq.1}) can be determined from a perturbative expansion of the
Casimir free energy in the deformation profile $h({\bf x})$, defined above the point of closest proximity as $H({\bf x})=d+h({\bf x})$, see Fig.~\ref{fig1}.  Note that the latter perturbation requires a small deformation amplitude while the gradient expansion relies upon small changes in the slope.
However, it can be shown that under certain conditions (existence of perturbation theory in $h({\bf x})$ and regularity of the involved kernels in momentum space) the gradient expansion of Eq.~(\ref{eq.1}) follows from resumming the perturbative series for small in-plane momenta.
If so, the expansion of the free energy in Eq.~(\ref{eq.1}) has to match  the perturbative expansion
for $|h({\bf x})|/d\ll 1$ {\em at small momenta}. The latter is given by
\begin{align}
{\cal F}[d+h({\bf x})]&=A\, {\cal F}_{pp}(d)+\,\mu(d)  {\tilde h}({\bf 0}) \nonumber \\
&+ \,\int \frac{d^2{\bf k}}{(2 \pi)^2} \,{\tilde G}({ k};d)\,|{\tilde
h}({\bf k})|^2+{\bar \rho}^{(2)}[h]\;,\label{eq.2}
\end{align}
where $A$ is the surface area,  ${\bf k}$ is the in-plane wave-vector,
${\tilde h}({\bf k})$ is the Fourier transform of $h({\bf x})$,
and ${\bar  \rho}^{(2)}[h]$ refers to higher order corrections.
The function $\alpha(H)$ can now be determined {\em if} the kernel ${\tilde G}({ k};d)$ can
be expanded to second order in $k$ (which as we shall see is the case for our problem).
Indeed, matching the expansion
\begin{equation}
{\tilde G}({k};d)=\gamma(d)+\delta(d)\,k^2 \,+o(k^2)\;,
\label{eq.3}
\end{equation}
to Eq.~(\ref{eq.1}) leads to
 \begin{equation} {\cal F}_{pp}'(d)= \mu(d) \, ,\quad {\cal F}_{pp}''(d)=2 \gamma(d) \, ,
 \quad \alpha(d)= \delta(d) \;,
\label{eq.6}
\end{equation}
where a prime denotes a derivative with respect to $d$.  For further justification of the gradient expansion we refer to the discussion after Eq.~\eqref{eq.A}.

We now give an outline of the computation of the kernel ${\tilde G}({k};d)$. The staring point is the
scattering formula for  the Casimir free energy ${\cal F}$,~\cite{universal}
\begin{equation}
\label{eq.4}
{\cal F}= k_B T {\sum_{n \ge 0}}' \text{Tr} \log [1- {\mathbb T}^{(1)} {\mathbb U} {\mathbb T}^{(2)}  {\mathbb U}  ]\;,
\end{equation}
where $k_B$ is Boltzmann's constant, $T$ the temperature, and the
primed sum runs over Matsubara frequencies $\xi_n= 2 \pi n k_B T
/\hbar$ with the $n=0$ term weighted by $1/2$.  In Eq.~(\ref{eq.4}),
${\mathbb T}^{(j)}$ denotes the T-operator of plate $j$, evaluated for
imaginary frequency $i\xi_n$.  In a plane-wave basis $|{\bf k},Q
\rangle$,\cite{fn3} where $Q=E,M$ is the polarization, $E$ and $M$
denote respectively electric (transverse magnetic) and magnetic
(transverse electric) modes.  The translation operator $\mathbb U$ in
Eq.~(\ref{eq.4}) is diagonal with matrix elements $e^{-d q_n}$ where
$q_n=\sqrt{{ k}^2+ \kappa_n^2}\equiv q_n(k)$, $k=|{\bf k}|$ and
$\kappa_n=\xi_n/c$.  For the undeformed slab, the operator ${\mathbb
  T}^{(2)} $ is diagonal in the plane-wave basis with matrix elements
given by the scattering amplitudes \be { {\cal T}}_{QQ'}^{(2)}({\bf
  k}, {\bf k}' ) =(2 \pi)^2 \delta^{(2)}({\bf k}-{\bf
  k'})\,\delta_{QQ'}\, r^{(2)}_{Q} (i \xi_n,{\bf k})\;, \ee where
$r_{Q}^{(j)} (i \xi_n,{\bf k})$ are the Fresnel reflection
coefficients
\begin{equation}
r_{E}^{(j)}  (i \xi_n,{k})=\frac{\epsilon_j(i \xi_n)\,q_n-s_n^{(j)}}{\epsilon_j(i \xi_n)\,q_n+s_n^{(j)}}\, ,
r_{M}^{(j)}  (i \xi_n,{ k})=\frac{ q_n-s_n^{(j)}}{ q_n+s_n^{(j)}}\,,
\end{equation}
and $s_n^{(j)}=\sqrt{\epsilon_j( i\,\xi_n) \kappa_n^2+{ k}^2}$.
There are no analytical formulae for the elements of the T-operator of the curved plate ${\mathbb T}^{(1)} $,
but for small  deformations they can be expanded in powers of  $h({\bf x})$ as\cite{voron}
\begin{align}
& { {\cal T}}_{QQ'}^{(1)}({\bf k}, {\bf k}' )=(2 \pi)^2 \delta^{(2)}({\bf k}-{\bf k'})\,\delta_{QQ'}\, r^{(1)}_{Q} (i \xi_n,{\bf k})
\nonumber\\
&+ \sqrt{q_n\,q'_n}\,\left[-2 \,B_{QQ'}({\bf k}, {\bf k}')\,\tilde{h}({{\bf k}- {\bf k}'})\right. \\
 & \left. + \!\!\int \!\!\frac{d^2 {\bf k}''}{(2 \pi)^2} (B_2)_{QQ'}({\bf k}, {\bf k}';{\bf k}'') \tilde{h}({{\bf k}\!- \!{\bf k}''}) \tilde{h}({{\bf k}''\!-\! {\bf k}'})+\dots \right]\;, \nonumber
\end{align}
where $q'_n=q_n(k')$. The coefficients $B_{Q Q'}({\bf k}, {\bf k}')$ and $(B_2)_{Q Q}({\bf
  k}', {\bf k}';{\bf k}'')$ can be obtained by standard perturbation
theory in the height field and are given in Ref.~\onlinecite{voron}; they depend on
the relative orientation of the wave vectors ${\bf k}$ and ${\bf k}'$, on the
corresponding $q_n$ and $s_n$ and the dielectric function itself.
Substituting the above expansion 
into Eq.~(\ref{eq.4}), we obtain
\begin{equation}
{\tilde G}({k};d)={k_B T} \!{\sum_{n \ge 0}}'\!\!\int \!\!\frac{d^2 {\bf k}'}{(2 \pi)^2}\, \frac{f_n({\bf k'}, {\bf k}'\!+{\bf k})\!+\!f_n({\bf k}', {\bf k}'\!-{\bf k})}{2}\;,
\label{eq.7}
\end{equation}
where
\begin{align}
\label{eq.5}
& f_n({\bf k}', {\bf k}'')  =-  \sum_{Q}   \frac{q'_nr_{Q}^{(2)}(k')}{g_{Q}(k')}e^{-2 q_n' d}
\, [(B_2)_{QQ}({\bf k}', {\bf k}';{\bf k}'') \nonumber \\
& + 2 \sum_{Q'}\,\frac{q_n''\, r_{Q'}^{(2)}(k'')}{g_{Q'}(k'')}\,e^{-2 q''_n d}
B_{QQ'}({\bf k}',{\bf k}'') B_{Q' Q}({\bf k}'',{\bf k}')] \;,
\end{align}
with $q''_n=q_n(k'')$,   $g_{Q}(k)=1-r^{(1)}_{Q} r^{(2)}_{Q} \exp[-2 q_n d]$. The  explicit  dependence of several quantities on $i \xi_n$   is not shown for brevity.
 Since the sum over Matsubara frequencies in Eq.~(\ref{eq.7}) is exponentially convergent,  the existence of the second $k$-derivative of $\tilde G(k;d)$ is ensured if for all $n \ge 0$ the second derivative of   $f_n({\bf k}', {\bf k}'+{\bf k})$ with respect to say  $k_x$, is absolutely integrable over ${\bf k}'$ for ${\bf k}={\bf 0}$. We then have
\begin{equation}
\label{eq.A}
\alpha(d)=\left.\frac{1}{2}\frac{\partial^2 {\tilde G}}{\partial k^2}\right\vert_{k=0}\!\!\!\!\!\!=\left.{\frac{k_B T}{2}} {\sum_{n \ge 0}}'\!\!\int \!\!\frac{d^2 {\bf k}'}{(2 \pi)^2} \frac{\partial^2 f_n ({\bf k}', {\bf k}'\!\!+{\bf k})}{\partial k_x^2}\right\vert_{{\bf k}={\bf 0}} . 
\end{equation}

Having obtained the amplitude function $\alpha(H)$  by a comparison
with the perturbative free energy at small in-plane momenta, we note
that in principle, the free energy of Eq.~\eqref{eq.1} should follow
directly from Eq.~\eqref{eq.4} when the T-operators are expanded in
powers of the {\it gradient} of the surface profile $h({\bf x})$,
making no assumption about the {\it amplitude} of $h({\bf x})$
itself. Indeed, such an expansion has been carried out for dielectric
surfaces  in Ref.~\onlinecite{voron}. Substitution of this gradient expansion in Eq.~\eqref{eq.4} yields a {\it non-local} functional for the free energy.
Hence, the function $\alpha(H)$ must be determined by the gradient expansion of the T-operator and subsequent locality expansion of the free energy.\cite{next}

The above formulae enable evaluation of $\alpha(d)$ for arbitrary
dielectric functions $\epsilon_1(\omega)$ and $\epsilon_2(\omega)$. In
the ideal limit of perfect conductors at zero temperature (as well as
for scalar fields obeying Dirichlet, Neumann and mixed boundary
conditions), $\alpha(d)$ has a simple power law dependence on $d$; as
$\alpha(d) \sim 1/d^3$ with a coefficient that can be computed
exactly. This permits to obtain simple closed formulae for the leading
correction to the PFA for a variety of profiles $h({\bf
  x})$.\cite{Bimonte2011} In general, for finite $T$ and/or for
dielectric materials, the dependence of $\alpha(d)$ on $d$ cannot be
expressed as a simple power law, and has to be computed numerically.
Evaluation of Eq.~(\ref{eq.1}) then permits to estimate the leading
correction to PFA for arbitrary shapes of the profiles, and for any
materials and temperatures of the involved bodies.

While numerical evaluation of Eq.~(\ref{eq.1}) for specific
experimental setups (e.g. spheres, cylinders or corrugated plates) is
straightforward in general, by simple manipulations of
Eq.~(\ref{eq.1}) not presented here for brevity \cite{next}, it is
possible to obtain a closed semi-analytical expression for the leading
correction to PFA for the {\it gradient} of the Casimir force
$\partial F/ \partial d=-\partial^2 {\cal F}/\partial d^2$ (a quantity
that is measured directly in some experiments\cite{decca}), in the
experimentally relevant geometry of an arbitrary but axially symmetric
profile $h({\bf x})$, which is a smooth function of
$\rho^2=x_1^2+x_2^2$.  We expand the profile around the point of
minimum separation as
\begin{equation}
h(x_1,x_2)\equiv h(\rho^2)=d+\frac{\rho^2}{2 R}+c_1\frac{\rho^4}{2\,R^3} + \dots\;,
\label{prof}
\end{equation}
where $R$ sets the overall scale of curvature. By substituting the above expansion into Eq.~(\ref{eq.1}), one can derive the leading correction to PFA for the force-gradient as
\begin{equation}
\frac{\partial F}{\partial d}=- {2\pi R} F_{pp}(d)\left(1+\hat{\theta}_1\,\frac{d}{R} + o ( {d}/{R} )
 \right)\;,
\label{eq.8}
\end{equation}
where $F_{pp}(d)=-\partial  {\cal F}_{pp}(d)/ \partial d$ is the Casimir force per unit area between two parallel plates,   and
\begin{equation}
\hat{\theta}_1=  \frac{{\cal F}_{pp}(d)}{d\, F_{pp}(d)} \left( 2 \beta(d)- 4\, {c_1}  \right), \, \quad
\beta(d)=\frac{\delta (d)}{{\cal F}_{pp}(d)} \, .
\label{eq.9}
\end{equation}
Note that the dimensionless function $\beta(d)$ depends on the ratio of $d$ to one (or more) material dependent length scale.  Independently of the dielectric material, the function $\delta(d)$ becomes a simple power law in the non-retarded limit where ${\cal F}_{pp}(d)\sim 1/d^2$ and $\delta(d)\sim1/d^2$ so that $\beta(d)$ and hence $\hat \theta_1$ tends to a constant for $d\to 0$. This implies that in the non-retarded limit the first correction to the PFA for ${\cal F}$  scales logarithmically with the separation $d$.

\begin{figure}
\includegraphics[width=1.\columnwidth]{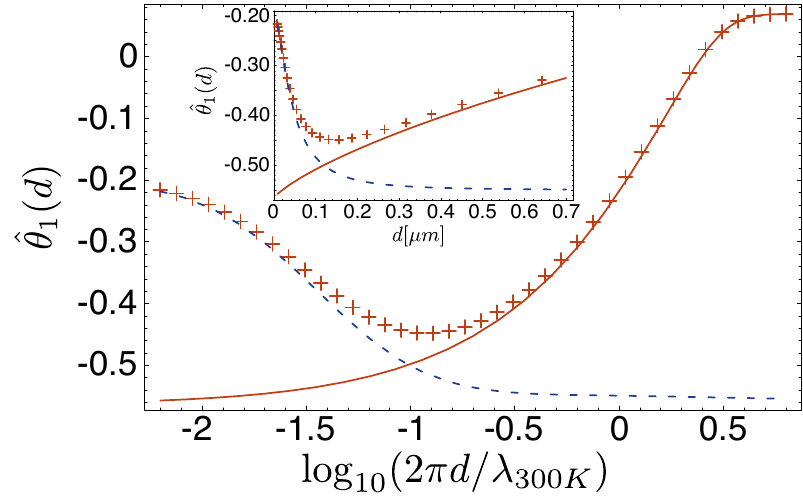}
\caption{\label{fig2} $\hat{\theta}_1$   for a gold sphere in front of
  a gold plate,  computed using the tabulated optical data for gold,
  see Ref.~\onlinecite{palik}. Crosses correspond to $T=300$~K, while the dashed line
is for  $T=0$~K.  The solid line is for an  ideal conductor at $T=300$~K. For ideal conductors at $T=0$,
$\hat{\theta}_1=-0.564$ independently of separation.
The inset depicts the same data, for gold at $300$~K, as a function of separation (in microns)
on a linear scale.}
\end{figure}

Using the above formulae, we numerically computed the coefficient
$\hat{\theta}_1$ in the experimentally relevant sphere-plate geometry
(corresponding to $c_1=1/4$), assuming for simplicity that both
surfaces are made of the same material; gold at
$T=300$~K.  For a perfect reflector at zero temperature, an exact
value for $\hat{\theta}_1$ can be obtained. In this limit the
coefficient $\beta$ was computed in Ref.~\onlinecite{Bimonte2011}, and
given as $\beta_{\rm EM}=2/3 (1-15/\pi^2)$.  Since for perfect
reflectors at $T=0$, ${\cal F}_{pp}=-\pi^2 \hbar c/(720 d^3)$, it
follows from Eq.~(\ref{eq.9}) that
$\hat{\theta}_1\vert_{T=0}^{\epsilon=\infty}= \frac{1}{3}\left( 2
  \beta_{\rm EM} - 1 \right) = -0.564 $.  Away from this ideal limit,
$\hat{\theta}_1$ has to be computed numerically. In our computations
for gold, we adopted the Drude model (for a discussion of alternative
models see Ref.~\onlinecite{BKMM-book}, Chap. 13).  In this model, the
permittivity $\epsilon(i \xi)$ to be used in the computation of the
Casimir force for (ohmic) conductors is obtained using the well known
Kramers-Kronig relation.  For ${\rm Im}[ \epsilon(\omega)]$ we used
the tabulated optical data for gold quoted in Ref.~\onlinecite{palik},
extrapolated to zero frequency by means of the Drude model, with
parameters $\Omega_p=9$ eV$/\hbar$ and $\gamma=35$ meV$/\hbar$ (see
Ref.~\onlinecite{BKMM-book}, p.336).  The resulting $\hat{\theta}_1$
is plotted in Fig.~\ref{fig2}, as a function of $\log_{10} (2 \pi
d/\lambda_{300\,{\rm K}})$, where $\lambda_T=\hbar c/(k_B T)$ is the
thermal wavelength ($\lambda_{300\,{\rm K}}=7.6 \;\mu$m). The inset of
Fig.~\ref{fig2} shows these data versus the separation in linear scale
(in microns). The crosses in Fig.~\ref{fig2} correspond to data for
$T=300$~K, while the dashed lines are for $T=0$~K. We also show (solid
lines) the result for perfect reflectors at $T=300$~K. Thus, the solid
lines neglect finite conductivity corrections, while the dashed lines
neglect finite temperature corrections.  The importance of including both corrections for the leading (PFA)
Casimir force has been previously noted \cite{Canaguier2}. A comparable previous estimate of $\hat\theta_1$, for a
metal sphere-plate setup at zero temperature, was obtained in
Ref.~\onlinecite{Canaguier}. This estimate is based on a numerical summation of a partial
wave series, valid at large separations $d \gg R$, and fitting the results
at short separations $d \to 0$ to a polynomial series. While our explicit
computation here confirms a linear correction to PFA as in Eq.~\eqref{eq.8}, we
expect higher order corrections to involve non-polynominal logarithmic
terms\cite{Bimonte2011}.

Some comments about the qualitative behavior of the numerical results are in order.
As expected, the $T=0$ perfect conductor limit of -0.564 is approached by the perfect conductor data
at $T=300$~K (solid lines) for short separations, which is where temperature effects become negligible;
while the $T=0$ data for gold (dashed lines) approach this limit for large separations,
which is where the skin depth becomes negligible.
We observe also that in the limit of large separations, the $T=300$~K data in Fig.~\ref{fig2} (crosses and the solid line), both approach the limit $\hat{\theta}_1(d\to\infty)=1/(12 \,\zeta(3))=0.0693$, which coincides with  the classical high-temperature limit of a scalar field obeying Dirichlet boundary conditions. This is plausible since for finite $T$ in the limit of large separations the Casimir force is dominated by classical thermal fluctuations of the electromagnetic field.  It is well known that for non-magnetic materials like ohmic conductors, transverse electromagnetic fields  decouple in the classical limit (Bohr-van Leeuwen theorem \cite{Bimonte2009}), and the Casimir force arises entirely from low-frequency fluctuations of the longitudinal electric field (non-retarded limit of E modes), associated with scalar potentials which vanish on the surface of a conductor.

Importantly, we see from Fig.~\ref{fig2} that the data for gold at $T=300$~K deviate significantly from both the solid and the dashed curves for separations $d$ between $150$nm and $300$nm, showing that in this range a precise determination of $\hat{\theta}_1$ requires {\it simultaneous} consideration of both finite conductivity {\it and} finite $T$ corrections. This is somewhat surprising since the thermal correction to the Casimir force between two metallic plates is small for such small separations (in the Drude model, less than $3\%$ below $200$nm).  On the contrary, we see from Fig.~\ref{fig2} that the magnitude of the thermal correction to $\hat{\theta}_1$ is about $20\%$ for $d=200$nm, indicating that $\hat{\theta}_1$ is much more sensitive to temperature than the Casimir force itself.  For a vanishing separation we get $\hat \theta_1(d\to 0)=-0.206$ which is consistent with our general prediction of a finite $\hat \theta_1$ in the non-retarded limit, and suggests that $\hat\theta_1(d)$ is a bounded function.

Since Casimir force measurements are often limited to rather short separations between weakly curved surfaces, our work has immediate experimental relevance.  An attempt to measure the coefficient ${\hat \theta}_1$ for a gold plate-sphere geometry at room temperature was made by using a micromachined torsional oscillator.\cite{deccaonPFA} The best estimate of ${\hat \theta}_1$ was obtained by dynamically measuring the Casimir force gradient at fixed separation $d$, for five spheres of different radii $R$, ranging from $10.5$ to $148.2\mu$m, and fitting the data linearly in $1/R$.  It was estimated that ${\hat \theta}_1$ is less than $0.4$ at 95$\%$ confidence level, in the range $164$nm $< d < 300$nm. Our data in Fig.~\ref{fig2} are  slightly above, but roughly consistent with this bound. The small disagreement between our estimate of ${\hat \theta}_1$ and the experimental bound may be explained by observing that the theoretical prediction of ${\hat \theta}_1$ is sensitive to the plasma frequency $\omega_p$, with smaller $\omega_p$ leading to smaller values for ${\hat \theta}_1$. In our computations we used the standard value for gold ($\omega_p=9$ eV/$\hbar$), but it is well known that gold films produced by deposition techniques may have much smaller values of the plasma frequency.\cite{sveto}
Another consideration is that, when fitting the force-gradient data  in Ref.~\onlinecite{deccaonPFA} higher order corrections,  of magnitude $\hat{\theta}_2(d/R)^2$, to  PFA were neglected.  The coefficient $\hat{\theta}_2$ cannot be computed analytically yet, but its magnitude was estimated for perfect conductors at zero temperature through a Pad\'e extrapolation constrained by a multipole expansion at large separations and by the gradient expansion at short separation.\cite{Bimonte2011}
It appears that $\hat{\theta}_2$ is positive and of order one. Neglect
of this second order correction in the  fits of
Ref.~\onlinecite{deccaonPFA}, especially for the smallest sphere, may
lead to a systematic underestimation of  ${\hat \theta}_1$. Finally,
it would be important to investigate the influence of surface roughness on $\hat{\theta}_1$.

This research was supported by the ESF Research Network CASIMIR (GB, TE),
and NSF Grant No.~DMR-08-03315 (MK).

\end{document}